\documentclass[paper]{ieice}
\usepackage{pstricks,pst-node}
\usepackage{latexsym}

\newcommand{\startchange}{}
\newcommand{\finishchange}{}
\newcommand{\change}[1]{{#1}}

\field{A}
\vol{90}
\no{9}
\SpecialSection{Information Theory and its Applications}
\title[Optimal Network Error-Correcting Codes]{Construction Algorithm for Network Error-Correcting Codes Attaining the Singleton Bound}
\titlenote{To appear in IEICE Trans.\ Fundamentals (http://ietfec.\\ oxfordjournals.org/), vol.\ E90-A, no.\ 9, Sept.\ 2007. The page number has not been determined yet.}
\authorlist{% fill arguments of \authorentry, otherwise error will be caused. 
 \authorentry[ryutaroh@rmatsumoto.org]{Ryutaroh MATSUMOTO}{m}{titech}
}
\affiliate[titech]{The author is with the Department of Communications and
Integrated Systems, Tokyo Institute of Technology, Tokyo, 152-8550 Japan.}
\received{2006}{12}{10}
\revised{2007}{3}{30}
\finalreceived{2007}{5}{9}

\newtheorem{lemma}{Lemma}
\newtheorem{example}[lemma]{Example}
\newtheorem{proposition}[lemma]{Proposition}
\newtheorem{definition}[lemma]{Definition}

\newtheorem{remark}[lemma]{Remark}

\newcommand{\startproof}{\noindent\emph{Proof.} }
\providecommand{\qed}{\QED}

\begin{document}
\maketitle
\begin{summary}
We give \change{a centralized deterministic}
algorithm for constructing linear network
error-correcting codes that attain
the Singleton bound \change{of} network error-correcting codes.
The proposed algorithm is based on the algorithm by
Jaggi et~al.
\change{We give estimates on the time complexity and the
required symbol size of the proposed algorithm.
We also estimate the probability of
a random choice of local encoding vectors by all intermediate
nodes giving a network error-correcting codes attaining
the Singleton bound.}
We also clarify the relationship between
the robust network coding and the network error-correcting
codes with known locations of errors.
\end{summary}

\section{Introduction}
Ahlswede et~al.\ \cite{ahlswede00}
proposed the notion of network coding
that multicasts data from a single sender to
multiple receivers at a rate at which
the ordinary store and forward routing cannot
multicast the data.
Such high rate multicast becomes feasible by allowing
intermediate nodes to encode and decode the data.
A sender is usually called a source and a receiver is called
a sink.
A network coding is said to be linear if
every intermediate node outputs a linear combination
of its inputs \cite{li03}.

A study of network coding usually assumes that
an error does not occur in networks.
Recently, Cai and Yeung  \cite{yeung06a,yeung06b}
considered errors in network coding,
and proposed the network error correcting codes
that allow sinks to recover the information even
when errors occur on intermediate edges in the network.
After formulating the network error correction,
they proposed the lower and upper bounds
on the number of messages in a network $\alpha$-error correcting
code, \change{and one of their upper bound was a
natural generalization of the Singleton bound
for the ordinary error-correcting codes.
Recently,
Zhang \cite{zhang06} and Yang et~al.~\cite{yang07b}
independently observed that the Singleton bound
can be refined.
We note that the problem formulation in \cite{yeung06a,yeung06b}
was later independently presented in \cite{jaggi05}.
(The proceedings paper of \cite{yeung06a,yeung06b} appeared
in 2002.)}

Cai and Yeung  mostly considered the case that
intermediate nodes perform only simple encoding and
decoding without delay, such as computing the output of the node
as a linear combination of its inputs, and the sinks
perform complex decoding computation.
The network error correcting codes
\change{can} avoid  introducing decoding computation and delay into
intermediate nodes, which is the advantage 
over use of ordinary error correcting codes
between nodes.

Note that
a similar type of network
failure in a slightly different context  was considered in
\cite[Sect.\  V]{koetter03}
and
\cite[Sect.\  VI]{sanders05}
in which
every sink is assumed to know the set of failed edges
and failed edges are assumed to emit zero symbols.
Network error correction does not assume
the knowledge of edges causing errors,
and the problem formulation is different from \cite{koetter03,sanders05}.
Note also that Kurihara \cite{kurihara06} considered the different
notion of robustness. In his paper, he considered network coding
that allows sinks to recover partial information with edge failures.

\startchange
For the construction of the network error-correcting
codes, Jaggi et~al.~\cite{jaggi07}
proposed a randomized construction that uses
coding among different time intervals.
Their method produces
codes attains the Singleton bound 
with high probability with sufficiently long
block length, where the block length
refers to the number of time intervals
among which coding is done.
It is desirable to have a network
error-correcting code that does not
code among different time intervals and
thus does not introduce delay.
Concurrently to this paper,
Yang et~al.~\cite{yang07b}
proposed an explicit construction algorithm
that produces codes attaining the
refined Singleton bound.
The idea in \cite{yang07b} is similar to this paper
in the sense that they also regard errors as information
from the source and add extra components in the global
encoding vectors corresponding to errors.
\finishchange

In this paper,
we give a \change{deterministic and
centralized} algorithm that constructs
a network error-correcting code that \change{attains}
the Singleton bound \change{of} network error-correcting codes
obtained in \cite{yeung06a}.
\change{We also give a relationship between
the success probability and the field size
for successful construction of network error-correcting
codes when intermediate nodes choose their encoding
coefficients randomly and independently.}
The proposed algorithms are based on \cite{sanders05}.
Our network error-correcting codes
make multicast robust to errors without introducing
delay in the transmission, which is very attractive to
delay sensitive multicast applications, such as
multicast of video or audio.
Our method is also useful for cryptographic applications,
because it can tolerate modification and deletion of data
by an adversary.

This paper is organized as follows.
Section 2 introduces notations and the model of errors.
Section 3 proposes an algorithm for constructing
network error-correcting codes \change{attaining}
the Singleton bound.
\change{Section 4 shows how to modify the algorithm
in Sect.\  3 to attain the refined Singleton bound,
the success probability of the random
construction of network error-correcting codes,
and the relationship between
the robust network coding \cite{koetter03,sanders05}
and the network error-correcting
codes with known locations of errors \cite{yang07}.}
Section 5 gives concluding remarks.

\section{Preliminary}
\subsection{Basic notations}
We consider an acyclic directed graph $G=(V,E)$ with possible parallel
edges of unit capacity.
$V \ni s$ denotes the source  and $V \supset T$ denotes
the set of sinks.
Let $n$ be the smallest min-cut separating $s$ from any $t\in T$
\change{throughout this paper.}
For $v \in V$, $\Gamma^+(v)$ (resp.\ $\Gamma^-(v)$) denotes the 
set of edges leaving (resp.\ reaching) the node $v$,
and $\mathrm{start}(e)$ (resp.\ $\mathrm{end}(e)$)
denotes the node at which the edge $e$ starts (resp.\ ends).

We consider linear coding over a finite field $\mathbf{F}_q$ with
$q$ elements.
The source  $s$ gets $k$ ($\leq n$) input symbols from $\mathbf{F}_q$.
The symbol $y(e)\in \mathbf{F}_q$ carried by an edge $e$ is a linear combination of
the symbols carried by the edges entering $\mathrm{start}(e)$.
The \emph{local \change{encoding} vector
$m_e: \Gamma^-(\mathrm{start}(e))
\rightarrow\mathbf{F}_q$}  determines the coefficients of
this linear combination, that is,
\[
y(e) = \sum_{e'\in \Gamma^-(\mathrm{start}(e))} m_e(e')y(e').
\]

In this paper,
a nonsink node performs only the computation of linear combination
of its inputs, and they do not correct errors.
An error is assumed to occur always at an edge.
When an error occurs at an edge $e$,
the symbol received by $\mathrm{end}(e)$ is different from
one sent by $\mathrm{start}(e)$,
and $\mathrm{end}(e)$ computes its outputs as if there was no error
at $e$. The error value at an edge $e$
is defined by the received symbol minus
the transmitted symbol at $e$.
Note that we express
a failure of a node $v \in V$ in a real network as
errors on edges in $\Gamma^+(v)$ in our model.
The number of errors is the number of edges at which errors occur.
A network code is said to correct $\alpha$ errors
if every sink can recover the original information
sent by the source when $\alpha$ or less
errors occur at arbitrary edges.
\change{We call the recovery of information by a sink \emph{decoding}.}

We represent errors occurred in the whole network
by a vector $\vec{e}$ in $\mathbf{F}_q^{|E|}$,
where $|E|$ denotes the number of elements in $E$.
Fix some total ordering in $E$, and enumeration of
the error values gives $\vec{e}$.

Regarding on the number of messages in a network $\alpha$-error
correcting code, Cai and Yeung obtained the following result.
\begin{proposition}\label{lemsingleton} \textnormal{\cite{yeung06a}}
The number $M$ of messages
in  a network $\alpha$-error
correcting code, not necessarily linear, is upper bounded by
\[
M \leq q^{n-2\alpha}.
\]
\end{proposition}

\startchange
Very recently, Zhang~\cite{zhang06} and
Yang et~al.~\cite{yang07b} observed that the above
proposition can be refined as follows.

\begin{proposition}\label{lemsingleton2} \textnormal{\cite{zhang06,yang07b}}
Let $n_t$ be the min-cut from the source $s$ to a sink $t$.
If the sink $t$ can correct any $\alpha_t$ errors then
the number $M$ of messages
in  the network 
correcting code, not necessarily linear, is upper bounded by
\[
M \leq q^{n_t-2\alpha_t}.
\]
\end{proposition}
\finishchange

\subsection{Jaggi et~al.'s algorithm for construction of
an ordinary network \change{code}}
In this subsection,
we review Jaggi et~al.'s algorithm \cite{sanders05} for construction of
an ordinary network coding.
The proposed algorithm uses a modified version of their algorithm.

Since linear coding is used,
the information carried by an edge $e$
is a linear combination of $k$ information symbols in $\mathbf{F}_q$.
We can characterize the effect of all the local
\change{encoding} vectors on an edge $e$ independently of a
concrete $k$ information symbols using \emph{global
\change{encoding} vectors $\vec{b}(e) \in \mathbf{F}_q^k$}.
When the information from the source is $\vec{i} \in \mathbf{F}_q^k$,
the transmitted symbol on an edge $e$ is equal to the inner product
of $\vec{i}$ and $\vec{b}(e)$.
\change{In order to decide the encoding at the source node $s$,
we have to introduce an imaginary source $s'$ and
$k$ edges of unit capacity from $s'$ to $s$.
We regard that $s'$ sends $k$ symbols to $s$ over
$k$ edges.}

We initially computes an \change{$s'$-$t$} flow $f^t$
of magnitude $k$ for each $t\in T$ and decomposes this flow
into $k$ edge disjoint paths from \change{$s'$} to $t$.
If an edge $e$ is on some flow path $W$ from \change{$s'$} to
$t$, let $f_\leftarrow^t(e)$ denote the predecessor edge of
the edge $e$ on the path $W$.
Jaggi et~al.'s algorithm steps through 
the nodes $v\in V$ in a topological order
induced by the directed graph $G$.
This ensures that the global \change{encoding} vectors of all
edges reaching $v$ are known when the local \change{encoding} vectors
of the edges leaving $v$ are determined.
The algorithm defines the coefficients of $m_e$
for one edge $e\in\Gamma^+(v)$ after the other.
There might be multiple flow paths to different sinks
through an edge $e$. Let $T(e)$ denote the set of sinks
using $e$ in some flow $f^t$ and
let $P(e) = \{f_\leftarrow^t(e)
\mid t\in T(e)\}$ denote the set of
predecessors edges of $e$ in some flow path.
The value $0$ is chosen for $m_e(e')$ with edges
$e' \notin P(e)$.

We introduce two algorithmic variables $B_t$ and $C_t$
that are updated by Jaggi et~al.'s algorithm.
$C_t$ contains one edge from each path in $f^t$,
namely the edge whose global \change{encoding}
vector was defined most recently in the path.
$B_t = \{\vec{b}(e) \mid e \in C_t\}$ is updated
when $C_t$ is updated.
The algorithm determines $m_e$ so that
for all $t \in T$,
$B_t$ 
is linearly independent.

After finishing the algorithm,
every sink can \change{decode} the original information because $B_t$ is
linearly independent.

\section{Construction algorithm}\label{sec:const}
We shall propose an algorithm constructing a network $\alpha$-error
correcting code carrying $k$ information
symbols in $\mathbf{F}_q$ with $n-k\geq 2\alpha$,
which is equivalent to the Singleton bound (Proposition \ref{lemsingleton}).
The proposed construction is based on \cite{sanders05}.
We assume that the size of alphabet $\mathbf{F}_q$ satisfies
\begin{equation}
q > |T| \cdot {|E| \choose 2\alpha}. \label{qassumption}
\end{equation}

\begin{definition}\label{def1}
For the original information $\vec{i} \in \mathbf{F}_q^k$ and
the error $\vec{e} \in \mathbf{F}_q^{|E|}$,
let \change{$\phi_t(\vec{i},\vec{e}) \in \mathbf{F}_q^{|\Gamma^-(t)|}$}
be the vector of symbols carried by the input edges to $t$.
\end{definition}

\begin{lemma}
If a sink $t$ can \change{decode} the original information $\vec{i}$
with any $2\alpha$ or less errors whose locations are known to the sink $t$,
then the sink $t$ can \change{decode} the original information with any
$\alpha$ or less errors without the knowledge of the error locations
under the assumption that the number of errors is $\leq \alpha$.
\end{lemma}
Note that errors with known locations are called erasures in
\cite{yang07} and the properties of erasures are also studied
in \cite{yang07}.

\startproof
Denote the Hamming weight of a vector $\vec{x}$ by $w(\vec{x})$.
The assumption of the lemma implies that
for any $\vec{i} \neq \vec{j}$ and $\vec{e}$ with $w(\vec{e})
\leq 2\alpha$ we have
\begin{equation}
\phi_t(\vec{i},\vec{e})\neq 
\phi_t(\vec{j},\vec{0}). \label{eq10}
\end{equation}
Equation (\ref{eq10}) implies that for any $\vec{i} \neq \vec{j}$ and
$\vec{e}_1$, $\vec{e}_2$ with $w(\vec{e}_1) \leq \alpha$ and
$w(\vec{e}_2)\leq \alpha$ we have
\[
\phi_t(\vec{i},\vec{e}_1)\neq 
\phi_t(\vec{j},\vec{e}_2),
\]
which guarantees that $t$ can \change{decode} the original information
under the assumption that the number of errors is $\leq \alpha$
by exhaustive search.
\qed

\startchange
\begin{remark}
The above lemma does not guarantee the existence of
an efficient decoding algorithm.
\end{remark}
\finishchange

Fix $F \subset E$ with $|F| = 2\alpha$.
We shall show how to construct a network error-correcting code that
allows every sink to \change{decode} the original information when the errors can
occur only at $F$.
\change{We call $F$ the \emph{error pattern}.
The following description is a condensed version of the
proposed algorithm, which is equivalent to the full description with $\mathcal{F} = \{ F \}$ in Fig.~\ref{fig2} on p.~\pageref{fig2}.}
\begin{enumerate}
\startchange
\item\label{step0} Add the imaginary source $s'$ and draw $k$ edges from
$s'$ to $s$.
\finishchange
\item\label{step1} Add an \change{imaginary} node $v$ at the midpoint of each $e \in F$ and add
an edge of unit capacity from \change{$s'$} to each $v$.
\item\label{step2} For each sink $t$, do the following:
\begin{enumerate}
\item\label{step2a} Draw as many edge disjoint paths
from \change{$s'$} to $t$ passing through the \change{imaginary} edges
added at Step \ref{step1}
as possible.
Let \change{$m_t^F (\leq 2\alpha)$} be the number of paths.
\item\label{step2b} Draw $k$ edge disjoint paths passing
through $s$ that are also edge disjoint from
the \change{$m_t^F$} paths drawn in the previous step.
\end{enumerate}
\item\label{finalstep} Execute the algorithm by Jaggi et~al.\ 
with $\sum_{t\in T}(k+m_t^F)$ edge disjoint paths constructed in Step \ref{step2}.
\end{enumerate}

\startchange
\begin{figure}[t!]
\psset{unit=0.1\linewidth}
\begin{pspicture}(0,0)(10,13)
%\newpsobject{showgrid}{psgrid}{subgriddiv=1,griddots=10,gridlabels=6pt}
%\showgrid
\rput(5,12){\circlenode[]{sd}{\large $s'$}}
\rput(5,10){\circlenode[]{s}{\large $s$}}
\rput(2,8){\circlenode[]{1}{\large $1$}}
\rput(4,8){\circlenode[]{2}{\large $2$}}
\rput(6,8){\circlenode[]{3}{\large $3$}}
\rput(8,8){\circlenode[]{4}{\large $4$}}

\rput(3,9.5){\dianode[]{A}{\large $A$}}
\rput(2.5,6.5){\dianode[]{B}{\large $B$}}

\rput(3,5){\circlenode[]{5}{\large $5$}}
\rput(5,5){\circlenode[]{6}{\large $6$}}
\rput(7,5){\circlenode[]{7}{\large $7$}}
\rput(3,3){\circlenode[]{8}{\large $8$}}
\rput(5,3){\circlenode[]{9}{\large $9$}}
\rput(7,3){\circlenode[]{10}{\large $10$}}
\rput(1,1){\circlenode[]{11}{\large $t_1$}}
\rput(9,1){\circlenode[]{12}{\large $t_2$}}

\ncline[linestyle=dashed]{->}{sd}{A}
\ncline[linestyle=dashed]{->}{sd}{B}

\ncarc[arcangle=15.000000]{->}{sd}{s}
\ncarc[arcangle=-15.000000]{->}{sd}{s}

\ncline{->}{s}{A}
\ncline{->}{A}{1}
\ncline{->}{s}{2}
\ncline{->}{s}{3}
\ncline{->}{s}{4}
\ncline{->}{1}{11}
\ncline{->}{4}{12}
\ncline{->}{1}{B}
\ncline{->}{B}{5}
\ncline{->}{2}{5}
\ncline{->}{2}{6}
\ncline{->}{3}{6}
\ncline{->}{3}{7}
\ncline{->}{4}{7}

\ncline{->}{5}{8}
\ncline{->}{6}{9}
\ncline{->}{7}{10}
\ncline{->}{8}{11}
\ncline{->}{9}{11}
\ncline{->}{10}{11}
\ncline{->}{8}{12}
\ncline{->}{9}{12}
\ncline{->}{10}{12}
\end{pspicture}
\startchange
\caption{Example of a network with \change{imaginary} nodes and edges.
Nodes $A$ and $B$ are the \change{imaginary} nodes added in Step~\ref{step1}
and the dashed lines from \change{$s'$} to $A$ and $B$ represent
the \change{imaginary} edges added in Step~\ref{step1}. See Example~\ref{ex1}
for explanation.}\label{fig1}
\finishchange
\end{figure}
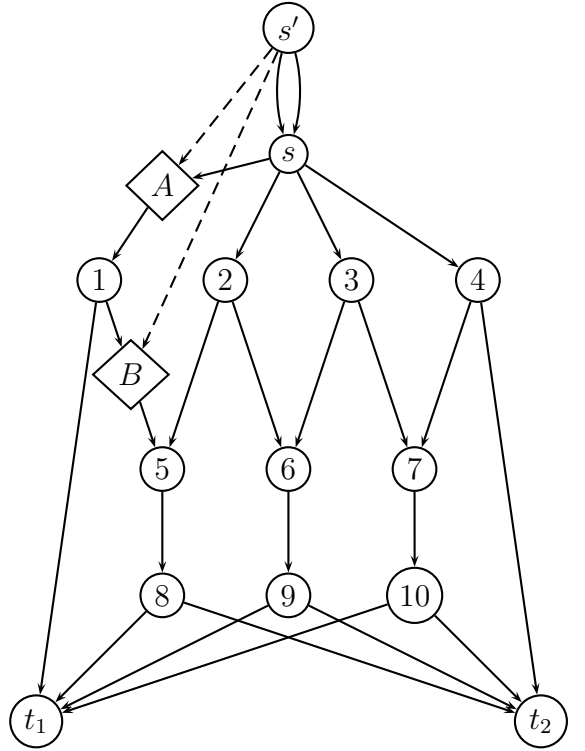

\begin{example}\label{ex1}
In Fig.~\ref{fig1},
we give an example of addition of \change{imaginary} nodes and edges.
The network structure in Fig.~\ref{fig1}
is taken from \cite[Fig.~2]{harada05}.
Nodes $A$ and $B$ are the \change{imaginary} nodes added in Step~\ref{step1}
and the dashed lines from \change{$s'$} to $A$ and $B$ represent
the \change{imaginary} edges added in Step~\ref{step1}.

The min-cut from $s$ to every sink is $4$ in the original network.
The set $F$ of edges with errors consists of the edge
from $s$ to node $1$ and the edge from node $1$ to node $5$.

We denote a path
by enumerating nodes on the path.
In Step~\ref{step2a} for $t_1$
we can find two edge disjoint paths,
namely $(s', A,1,t_1)$ and $(s',B,5,8,t_1)$.
On the other hand, in Step~\ref{step2a} for $t_2$,
we can find only one edge disjoint path,
namely $(s',A,1,B,5,8,t_2)$ \emph{or} $(s',B,5,8,t_2)$.
Therefore $m_{t_1}^F=2$ while $m_{t_2}^F = 1$.

In Step~\ref{step2b} for $t_1$,
we find two edge disjoint paths as
$(s',s,3,6,9,t_1)$ and $(s',s,4,7,10,t_1)$.
In Step~\ref{step2b} for $t_2$,
we find \emph{three} edge disjoint paths as
$(s',s,2,6,9,t_2)$, $(s',s,3,7,10,t_2)$, and $(s',s,4,t_2)$.
We can use arbitrary two paths among the three paths.
In either case, we can find $n-m_t^F$ paths in Step~\ref{step2b}. \qed
\end{example}

In Step~\ref{step2b}, we can guarantee the existence of
$k$ paths as follows:
Suppose that edges in \change{$m_t^F$} paths used in Step~\ref{step2a}
are removed from the original network $(V,E)$.
Then the min-cut from $s$ to a sink $t$ in the original network $(V,E)$
is at least $n - m_t^F$, which is larger than or equal to
$k$.

In Step \ref{finalstep}
we use the algorithm by Jaggi et~al.\  as if
the \change{imaginary} source
\change{$s'$} sent information on the $\alpha$ \change{imaginary} edges added
in Step \ref{step1}.
We denote by $B_t^F$ the set $B_t$ of global encoding vectors
for $k+m_t^F$ edge disjoint paths.
$B_t^F$ consists of $k+m_t^F$ vectors of length $k+2\alpha$.
We require that every sink $t$ is able to decode
$k$ information symbols, while $t$ may be unable to
decode $2\alpha$ error symbols in general
because $m_t^F \leq 2\alpha$.
\finishchange

There are always two edges end at the added \change{imaginary} node $v$
and one edge starts from $v$ in Step \ref{step1}. Since $v$ is \change{imaginary},
we cannot choose local \change{encoding} vectors at $v$.
Therefore, in Step \ref{finalstep},
all components in the local \change{encoding} vector at $v$ must be selected
to $1$, which keeps $B_t$ linearly independent.
\startchange
The reason is as follows:
Let $e$ be the edge from \change{$s'$} to $v$ added in Step~\ref{step1}.
The global \change{encoding} vector of $e$ is of the form
\[
(0^{j-1}, 1, 0^{n-j}),
\]
that is, it has only $1$ at the $j$-th component.
All other global \change{encoding} vectors in $B_t^F$ have zero
at the $j$-th component, since they are not in downstream
of $e$ when we choose local \change{encoding} vectors at $v$.
Therefore, the added \change{imaginary} node $v$
does not interfere with the execution of
Jaggi et~al.'s algorithm.
\finishchange

Observe also that $q > |T|$ guarantees the successful execution of
the algorithm as with the original version of Jaggi et~al.'s algorithm.

We shall show how each sink $t$ can \change{decode} the
original information sent from the source $s$.
\change{After executing Step~\ref{finalstep} we have decided all the
local \change{encoding} vectors in the original network $(V,E)$.}
Consider the three linear spaces defined by
\begin{eqnarray*}
V_1 &=& \{ \phi_t(\vec{i},\vec{e}) \mid \vec{i}\in \mathbf{F}_q^k,
\vec{e}\in\mathbf{F}_q^{|E|} \},\\
V_2 &=& \{ \phi_t(\vec{i},\vec{0}) \mid \vec{i}\in \mathbf{F}_q^k \},\\
V_3 &=& \{ \phi_t(\vec{0},\vec{e}) \mid \vec{e}\in\mathbf{F}_q^{|E|} \},
\end{eqnarray*}
where components in $\vec{e}$ corresponding to $E \setminus F$ are zero,
\startchange
and $\phi_t$ is as defined in Definition~\ref{def1}.
We consider $V_1$, $V_2$, and $V_3$ in the original network $(V,E)$
without added \change{imaginary} nodes and edges.
Then we have
\begin{equation}
V_1 = V_2 + V_3,\, \dim V_2 \leq k. \label{eq21}
\end{equation}
Since we keep $B_t^F$ linearly independent,
\begin{equation}
\dim V_1 \geq k+m_t^F. \label{eq22}
\end{equation}
Since the maximum number of edge disjoint paths passing through
the \change{imaginary} edges added in Step \ref{step1}
is \change{$m_t^F$}, we have
\begin{equation}
\dim V_3 \leq m_t^F. \label{eq23}
\end{equation}
Equations (\ref{eq21}--\ref{eq23}) imply
\begin{eqnarray}
\dim V_1 &=& k+m_t^F,\nonumber\\
\dim V_2 &=& k,\label{eq33}\\
\dim V_3 &=& m_t^F,\nonumber\\
\dim V_2 \cap V_3 &=& 0. \label{eq12}
\end{eqnarray}
The number of nonzero components in $\phi_t(\vec{i},\vec{e})$ is $k+m_t^F$
and the number of unknowns in $\phi_t(\vec{i},\vec{e})$ is $k+2\alpha$,
which can be larger than $k+m_t^F$.
However, by Eq.~(\ref{eq12}), the sink $t$ can compute
$\phi_t(\vec{i},\vec{0})$
from $\phi_t(\vec{i},\vec{e})$ as follows:
Write $\phi_t(\vec{i},\vec{e})$ as $\vec{u} + \vec{v}$
such that $\vec{u} \in V_2$ and $\vec{v} \in V_3$.
By Eq.~(\ref{eq12}) $\vec{u}$ and $\vec{v}$ are uniquely
determined \cite[p.19, Theorem 4.1]{lang87}.
We have $\vec{u} = \phi_t(\vec{i},\vec{0})$ and
the effect of errors is removed.
The sink $t$ can also compute the original
information $\vec{i}$ from $\phi_t(\vec{i},\vec{0})$
by Eq.~(\ref{eq33}).
\finishchange

We shall describe how to construct a network error-correcting
code that can correct errors in any edge set $F \subset E$ with
$|F| = 2\alpha$.
\change{Let $\mathcal{F} = \{ F \subset E \,:\, |F| = 2\alpha \}$.}
The idea in this paragraph is almost the same as the construction
of the robust network coding in \cite[Sect.\  VI]{sanders05}.
\change{Recall that}  $B_t^F$ is the set of global \change{encoding} vectors
on edge disjoint paths to a sink $t$ with an edge set $F$ of errors.
Execute Jaggi et~al.'s algorithm keeping $B_t^F$ linearly
independent for all $t\in T$ and all $F\in \mathcal{F}$. Then every sink $t$ can \change{decode} the original information
with the knowledge of the edge set $F$ on which errors actually occur.
As in \cite[Sect.\  VI]{sanders05},
\[
q > |T| \cdot |\mathcal{F}| = |T| {|E| \choose 2\alpha}
\]
guarantees the successful execution of the algorithm.

\startchange
We present a pseudo programming code
of  the proposed algorithm in Fig.\ \ref{fig2}.
In order to present a detailed description,
we introduce new notations.
$G_F=(V_F,E_F)$ denotes the network
with added \change{imaginary} nodes and edges in  Steps~\ref{step0}
and \ref{step1}
with the error pattern $F \subset E$.
Let $f^{t,F}$ be the flow established in Steps~\ref{step2a} and
\ref{step2b} in $G_F$.
Let $f_\leftarrow^{t,F}(e)$ denote the set of predecessor edges of
the edge $e$ in a flow path in $f^{t,F}$.
Let $T^F(e)$ denote the set of sinks
using $e$ in some flow $f^{t,F}$ and
let $P^F(e) = \{f_\leftarrow^t(e)
\mid t\in T(e)\}$.

\begin{figure}[t!]
\startchange
\begin{tabbing}
(* Initialization *)\\
Added \change{imaginary} node \change{$s'$} and edges $e_1$, \ldots, $e_{k}$
from \change{$s'$} to $s$. $O(k)$\\
%(* $e_1$, \ldots, $e_k$ carry the information *)\\
\textbf{foreach} error pattern $F \in \mathcal{F}$ \textbf{do}\\
\hspace*{4ex}\= Initialize global \change{encoding} vector\\
\>$\vec{b}^F(e_i) = (0^{i-1},1,0^{k+2\alpha-i})\in\mathbf{F}_q^{k+2\alpha}$.\`$O((k+2\alpha)^2)$\\
\> \textbf{foreach} edge  $e \in F$ \textbf{do}\\
\>\hspace*{4ex}\= Add an \change{imaginary} node $v$ at the midpoint of $e \in F$.\`$O(1)$\\
\>\> Divide $e$ into an edge to $v$ and an edge from $v$.\`$O(1)$\\
\>\> Draw an \change{imaginary} edge from \change{$s'$} to $v$.\` (*) $O(1)$\\
\>\textbf{endforeach}\\
\>\textbf{foreach} sink $t\in T$ \textbf{do}\\
\>\> Draw as many edge disjoint paths from \change{$s'$} to $t$ as possible\\
\>\>passing through the edge
added in (*).\\
\` $O(2\alpha(|E|+k+4\alpha))$\\
%\>\>Let \change{$m_t^F$} be the number of path.\\
\>\> Draw $k$ edge disjoint path from \change{$s'$} to $t$
passing through\\
\>\>$s$ and
also disjoint from
paths made in the previous step.\\
\` $O(k(|E|+k+4\alpha))$\\
\>\> Initialize the basis $B_t^F = \{ \vec{b}^F(e_i) \mid e_i$ is on a path to $t \}$.\\
\` $O((k+2\alpha)^2)$\\
\>\textbf{endforeach}\\
\textbf{endforeach}\\
(* Main loop *)\\
\textbf{foreach} edge $e\in \bigcup_{F\in\mathcal{F}} E_F
\setminus \{e_1, \ldots, e_k\}$ in a topological order \textbf{do}\\
\>\textbf{if} $\mathrm{start}(e) \in V$ \textbf{then}\\
\>\>Choose a linear combination $\vec{b}^F(e) = \sum_{p\in P^F(e)}m_e(p)\vec{b}(p)$\\
\>\>such that $B_t^F$ remains linearly independent for all $t$\\
\>\>and $F$ by the method in \cite[Sect.\  III.B]{sanders05}. \` (**)\\
\>\textbf{else}\\
\>\>$m_e(p) = 1$ for all $p\in P^F(e)$ and
$\vec{b}^F(e) = \sum_{p\in P^F(e)}\vec{b}(p).$\\
\` $O(k+2\alpha)$\\
\>\textbf{endif}\\
\textbf{endforeach}\\
\textbf{return} $\{ m_e(\cdot) \mid \mathrm{start}(e) \in V \}$.
\end{tabbing}
\caption{Construction algorithm for a network $\alpha$-error correcting code.
The rightmost $O(\cdot)$ indicates the time complexity executing
the step.}\label{fig2}
\finishchange
\end{figure}

We shall analyze the time complexity of the proposed
algorithm in Fig.~\ref{fig2}.
As in \cite{sanders05} we assume that any arithmetic
in the finite field is $O(1)$ regardless of the field size.
First we analyze that of the initialization part.
Observe that $|E_F| = |E| + k+ 2|F| = |E| + k+4\alpha$ because
each edge in $F$ adds two edges to $E$ and
there are $k$ edges from \change{$s'$} to $s$.
The most time consuming part in the initialization
is construction of edge disjoint paths,
whose overall time complexity
is $O((|E|+k+4\alpha)|\mathcal{F}||T|(k+2\alpha))$.

Next we analyze the time complexity of the main loop.
By \cite[Proof of Lemma 8]{sanders05},
the time complexity of choosing the local \change{encoding}
vector $m_e(p)$ in Step (**) is 
$O((|\mathcal{F}||T|)^2 (k+2\alpha))$,
which is the most time consuming part in the main loop.
Choice of $m_e(p)$
is executed for $|E|$ edges starting from a real node in $V$.
Thus, the time complexity of the main loop is
$O(|E|(|\mathcal{F}||T|)^2 (k+2\alpha))$,
and the overall time complexity is
$
O(
|\mathcal{F}||T|(k+2\alpha)[
|E|+k+4\alpha + |\mathcal{F}||T|])
$.
Note that $|\mathcal{F}| = {|E| \choose 2\alpha}$.

A sink \change{decode}s the information by exhaustive search.
Specifically the sink enumerates all the possible information
and all the possible errors for all $F\in\mathcal{F}$,
then compares the resulting symbols on incoming edges with
the actual received symbols by the sink.
The computation of the resulting symbols can be
done by a matrix multiplication in $O((k+\alpha)^2)$ time complexity.
The number of possible information is $q^k$ and
the number of possible errors is $\sum_{j=0}^{\alpha}{|E| \choose j}(q-1)^j$.
Thus, the time complexity of decoding by a sink is
$O(q^k \sum_{j=0}^{\alpha}{|E| \choose j}(q-1)^j(k+\alpha)^2)$.

\section{Variants of the proposed method and its relation to
the robust network coding}
We shall introduce two variants of the proposed method in this section.

\subsection{Attaining the refined Singleton bound}
Network error-correcting codes constructed by the proposed
method attains the Singleton bound (Proposition \ref{lemsingleton}),
while they do not necessarily attains the refined Singleton bound
(Proposition \ref{lemsingleton2}).
Yang et~al.\  \cite{yang07b} concurrently proposed a construction algorithm
that produces a code attaining the refined Singleton bound.
In this subsection we modify the proposed method so that
it can produce a code attaining the refined Singleton bound.

Let $n_t$ be the min-cut from $s$ to $t$,
and suppose that the source $s$ emits $k$ symbols within unit
time interval.
A sink $t$ can correct $\alpha$ errors if $2\alpha \leq n_t -k$.
Let $\mathcal{F}_t = \{ F\subset E \,:\, |F| = n_t - k\}$ and
$\mathcal{F} = \bigcup_{t\in T}\mathcal{F}_t$.
For fixed $F \in \mathcal{F}$ and $t\in T$,
we cannot garuantee that there exists $k$ edge disjoint
paths in Step~\ref{step2b}.
For such $F$, the sink $t$ cannot \change{decode} information
with errors occered at $F$. We exclude $B_t^F$ with such $(t,F)$
from the
algorithm.
Note that if $|F| \leq n_t - k$ then there always exist
$k$ edge disjoint paths in Step~\ref{step2b}.

In order to attain the refined Singleton bound
we keep the linear independence of
all bases in $\{B_t^F \mid t\in T$, $F\in\mathcal{F}$,
$|F| \leq n_t - k\}$ in Step (**) in Fig.~\ref{fig2}.
By the exactly same argument, we see that the produced code attains
the refined Singleton bound.

By almost the same argument as Sect.\ \ref{sec:const},
we see that the modified proposed algorithm runs
in time complexity 
$
O(
|\mathcal{F}||T|(k+2\alpha_\mathrm{max})[
|E|+k+4\alpha_\mathrm{max} + |\mathcal{F}||T|])
$,
where $\alpha_\mathrm{max} = \lfloor (\max_{t\in T} n_t - k)/2\rfloor$. The required field size for successful execution of the algorithm
is $|T| \cdot |\mathcal{F}|$, and in this case
$|\mathcal{F}|$ depends on the structure of the network $(V,E)$.

\begin{table*}[t!]
\startchange
\caption{Comparison among the proposed methods and \cite{yang07b,jaggi07}.
We assumed that the min-cut is $n$ for all $t\in T$
and $k = n-2\alpha$. $I$ denotes the maximum of in-degrees of nodes.}\label{tab1}
\begin{tabular}{|p{0.1\textwidth}|p{0.1\textwidth}|p{0.2\textwidth}|p{0.25\textwidth}|p{0.2\textwidth}|}\hline
&delay&required field size for the success probability of code construction to be
$\geq 1-\delta$&
time complexity of code construction&time complexity of decoding by sinks\\\hline\hline
Figure~\ref{fig2}&none&$|T| {|E| \choose 2\alpha}$&
 $O(
{|E| \choose 2\alpha}|T|(k+2\alpha)[
|E|+k+4\alpha + {|E| \choose 2\alpha}|T|])$&$O(q^k \sum_{j=0}^{\alpha}{|E| \choose j}(q-1)^j(k+\alpha)^2)$\\\hline
Sect.\ \ref{sec:random}&none&$|E||T|{|E| \choose 2\alpha}/\delta$&
$O(I)$&$O(q^k \sum_{j=0}^{\alpha}{|E| \choose j}(q-1)^j(k+\alpha)^2)$\\\hline
Paper \cite{yang07b}&none&$|T|{n + |E| -2 \choose 2\alpha}$
&
$O(|E| |T| q^k\sum_{j=0}^{2\alpha} {|E| \choose j}(q-1)^j)$&$O(q^k \sum_{j=0}^{\alpha}{|E| \choose j}(q-1)^j(k+\alpha)^2)$\\\hline
Paper \cite{jaggi07}&large&not estimated&$O(I)$&
$O((n \times \mathrm{delay})^3)$\\\hline
\end{tabular}
\finishchange
\end{table*}

On the other hand,
the time complexity of constructing
local \change{encoding} vectors by the method of Yang et~al.~\cite{yang07b}
is
\[
O\left( |E| q^k \sum_{t\in T} \sum_{j=0}^{n_t-k} {|E| \choose j}(q-1)^j\right),
\]
and the required field size is
\[
\sum_{t\in T} {n_t + |E| -2 \choose n_t-k}.
\]
The time complexity of the proposed algorithm
can be smaller or larger depending on the network structure and $q$
than Yang et~al.~\cite{yang07b}.
The required field size of the proposed algorithm
can also be smaller or larger depending on the network structure.
However, for the special case $n_t = n$ for all $t\in T$,
the required field size of the proposed method is
smaller than Yang et~al.~\cite{yang07b}.

\subsection{Completely randomized construction}\label{sec:random}
By using the idea in the previous section,
we can estimate the success probability of
constructing a network error-correcting code
by randomly choosing local \change{encoding} vectors
as follows.
The idea behind its proof is almost the same as
\cite[Theorem 12]{sanders05}.
Observe that the random choice of local \change{encoding} vectors
completely remove the time complexity
of selecting \change{encoding} vectors in the centralized manner
at the expense of larger required field size $q$.

\begin{proposition}
Suppose that the source $s$ transmits $k$ symbols within
unit time interval, and let $\mathcal{F} =\{F\subset  E \,:$
$|F| = 2\alpha\}$
be the set of edges on which errors can occur.
Suppose also that local \change{encoding} vector coefficients
are generated at random independently and uniformly over
$\mathbf{F}_q$.
With this network error-correcting code,
all sinks can correct errors in any edge set $F\in \mathcal{F}$
with probability at least $1-\delta$
if $q \geq |E||T||\mathcal{F}|/\delta$.
\end{proposition}

\startproof
First pick independent random local \change{encoding}
vectors for all edges in the network simultaneously.
Then pick an error pattern $F\in\mathcal{F}$.
For this $F$, execute Steps 1 and 2 in page~\pageref{step1}
and compute the global \change{encoding} vectors $\vec{b}^F(e)$'s belonging
to $\mathbf{F}_q^{k+|F|}$.
For each cut in the network,
test whether $B_t^F$'s are linearly independent for all $t$.
This test fails with probability at most $|T|/q$
by the proof of \cite[Theorem 9]{sanders05}
provided that this tests succeed on all the upstream cuts
and $n\geq k+2\alpha$.

In the proposed algorithm in Fig.~\ref{fig2},
we test linear independence of $B_t^F$'s on
$|E|$ cuts in Step (**), which is sufficient to
garuantee the decodability of the information by every sink.
By the same reason, for each sink to be able to correct errors in $F$,
one needs to consider linear independence only on
at most $|E|$ such cuts with random
choice of local \change{encoding} vectors.
By the union bound, the probability that
the the independence tests fails for any of $|T|$ sinks
in any of the $|E|$ cuts in any of the $|\mathcal{F}|$
error patters is at most $\delta$
if $q \geq |E||T||\mathcal{F}|/\delta$.
\qed

Jaggi et~al.~\cite{jaggi07} do not provide an estimate
on the relation between the success probability of
their algorithm and the field size $q$.
Their method \cite{jaggi07} uses coding among
different time intervals and thus introduces
delays while our methods do not introduce extra delay.
In addition to this,
$\alpha$-error correcting codes by constructed
by the proposed methods allow sinks to correct
less than $\alpha$ errors, while the method
in \cite{jaggi07} does not.
The advantage of the method in \cite{jaggi07}
over the proposed methods in this paper
is that their method allows efficient decoding of
information by every sink, while our proposed
methods require exhaustive search of transmitted
information.

We summarize the comparison among the proposed algorithms and
\cite{jaggi07,yang07b} in Table~\ref{tab1}.

\finishchange

\subsection{Relation to the robust network coding}
We clarify the difference between the robust network
coding in \cite[Sect.\  V]{koetter03},\cite[Sect.\  VI]{sanders05} and
the network error-correcting codes with known locations
of errors \cite{yang07}.
A network error correcting codes that can correct errors
on a known locations $F\subset E$
is a robust network coding tolerating 
edge failures on $F$.
However, the converse is not always true.
Consider the network consists of three nodes $\{s,t,v\}$
with two directed edges from $s$ to $v$ and
one directed edge from $v$ to $t$.
The source is $s$ and the sink is $t$.
The intermediate node $v$ sends to $t$ the sum of two inputs
from $s$. This network coding tolerate single edge failure
between $s$ and $v$ but cannot correct single error between
$s$ and $v$.

\section{Concluding remarks}
In this paper,
we proposed an algorithm constructing
network error-correcting
codes attaining the Singleton bound, and
clarified its relation to the robust network
coding \cite[Sect.\  VI]{sanders05}.

There are several research problems that have
not been addressed in this paper.
Firstly, the proposed deterministic algorithm requires
tests of linear independence against
${|E| \choose 2\alpha}$ sets consisting of \change{$k+m_t^F$} vectors,
which is really time consuming.
It is desirable to have a more efficient 
\change{deterministic} construction algorithm.

Secondly, since there seems no structure in the constructed
code, the \change{decoding} of the original information at a sink $t$
requires the exhaustive search by $t$
for possible information from the source and possible errors.
It is desirable to have a code with structure that allows
efficient decoding.

Finally, the case $|T|=1$ and $|E|=n$ includes the ordinary
error correcting codes as
a special case.
Substituting $|T|=1$, $|E|=n$ and $2\alpha = n-k$ into Eq.~(\ref{qassumption})
gives $q > {n \choose n-k}$,
which can be regarded as a sufficient condition
for
the existence of the MDS linear code.
On the other hand, a well-known sufficient condition for
the existence of the MDS linear code is $q > n-2$,
which suggests that Eq.~(\ref{qassumption})
is loose and that there is a room for improvement
in Eq.~(\ref{qassumption}).

\section*{Acknowledgment}
The author thanks for constructive criticisms by
reviewers that improved the presentation of the results
very much. He also thanks Dr.\ Masazumi Kurihara
for pointing out ambiguity in the earlier manuscript.
He would like to thank Prof.~Kaoru Kurosawa
for drawing his attention to the network error correction,
Prof.\ Olav Geil,
Prof.\ Toshiya Itoh, Prof.\ Tomohiko Uyematsu,
Mr.\ Akisato Kimura, and Dr.\ Shigeaki Kuzuoka 
for helpful discussions.
He also would like to thank Dr.\ Sidharth Jaggi and Mr.\ Allen Min Tan
for informing the papers \cite{jaggi07,yang07b}.
Part of this research was conducted during the author's
stay in the Department of Mathematical Sciences,
Aalborg University.

%\bibliographystyle{IEEEtran}
%\bibliographystyle{abbrv}
%\bibliographystyle{ieicetr}
%\bibliography{mrabbrev,mybib}

%\bibliographystyle{ieicetr}% bib style
%\bibliography{}% your bib database

%\begin{thebibliography}{99}% more than 9 --> 99 / less than 10 --> 9
%\bibitem{}
%\end{thebibliography}

%\profile{Ryutaroh MATSUMOTO}{was born in Nagoya, Japan, on November 29,
%1973. He received the B.E.\ degree in computer science, the M.E.\ degree
%in information processing, and the Ph.D.\ degree in electrical and
%electronic engineering, all from Tokyo Institute of Technology, Japan,
%in 1996, 1998, 2001, respectively.
%
%He was an Assistant Professor from 2001 to 2004, and has been an
%Associate Professor since 2004 in the Department of Communications and
%Integrated Systems of Tokyo Institute of Technology.
%His research interest includes error-correcting codes,
%quantum information theory, and communication theory.
%
%Dr.~Matsumoto received the Young Engineer Award and the Excellent
%Paper Award from IEICE as well as the Ericsson Young Scientist Award
%from Ericsson Japan in 2001.}
%\profile*{}{}% without picture of author's face
\end{document}